\documentclass[pra,showpacs,twocolumn]{revtex4}
\usepackage{bm,graphicx,amsmath}
\begin{document}
\title{Anomalous decoherence and absence of thermalization in a photonic many-body system}
\author{Jonas Larson}
\email{jolarson@fysik.su.se}
\affiliation{Department of Physics, Stockholm University,
AlbaNova University Center, SE-106 91 Stockholm, Sweden}

\date{\today}

\begin{abstract}
The intention of this work is twofold, first to present a most simple system capable of simulating the intrinsic bosonic Josephson effect with photons, and second to study various outcomes deriving from inherent or external decoherence. A qubit induces an effective coupling between two externally pumped cavity modes. Without cavity losses and in the dispersive regime, intrinsic Josephson oscillations of photons between the two modes occurs. In this case, contrary to regular Markovian decoherence, the qubit purity shows a Gaussian decay and recurrence of its coherence. Due to intrinsic non-linearities, both the Josephson oscillations as well as the qubit properties display a rich collapse-revival structure, where, however, the complexity of the qubit evolution is in some sense stronger. The qubit as a meter of the photon dynamics is considered, and it is shown that qubit dephasing, originating for example from non-demolition measurements, results in an exponential destruction of the oscillations which manifests the collectiveness of the Josephson effect. Non-selective qubit measurements, on the other hand, render a Zeno effect seen in a slowing down of the Josephson oscillations. Contrary to dephasing, cavity dissipation results in a Gaussian decay of the scaled Josephson oscillations. Finally, following Ponomarev {\it et al.} [Phys. Rev. Lett. {\bf 106}, 010405 (2011)] we analyze aspects of thermalization. In particular, despite similarities with the generic model studied by Ponomarev {\it et al.}, our system does not seem to thermalize.    
\end{abstract}
\maketitle

\section{Introduction}
Cavity quantum electrodynamics (QED) systems, and later circuit QED, operating in the strong coupling regime have proven to be excellent for studies of quantum phenomena such as entanglement/non-locality and non-classical states of the radiation field~\cite{haroche,cavityqed}. In the recent past, however, new directions in these fields have been explored. A long term goal is to build many-body quantum models constructed of interacting photons. The idea is that these systems can simulate many-body models appearing in fields such as condensed matter theory~\cite{hartman}. By now, several proposals of such cavity/circuit QED quantum simulators have been put forward, among others; Mott insulator--superfluid phase transitions~\cite{jcbh}, spin-models and fractional quantum Hall states~\cite{spin}, the anomalous Hall effect~\cite{AHE}, relativistic effects~\cite{klein}, non-Abelian models~\cite{nonAb}, the Jahn-Teller effect~\cite{jonasjt}, and the photonic Josephson effect~\cite{photJJ1,photJJ2}. The desired non-linearity, and thereby photon-photon interaction, plays an important role in all of these and is established via interaction between light and matter. In the simplest example of matter-light interaction, the Jaynes-Cummings model (JC)~\cite{jc,jonasjc}, the interaction is inherently non-linear giving rise to the effective scattering between photons. This far, for the JC model the most striking outings of the non-linearity are the collapse-revival~\cite{cr} and the photon blockade effects~\cite{blockade}. 

Experimentally, quantum simulators involving many-body systems are mainly to be found within ultracold atomic gases, either trapped dilute fermi or bose gases~\cite{pethick} or more strongly correlated lattice or low dimensional models~\cite{maciek}. These systems provide an extreme controllability of system parameters as well as being well isolated from their environments and hence possessing long coherence times. Together with the experimental achievements within this area, topics such as decoherence and thermalization in systems processing many degrees-of-freedom have gained extra attention recently. 

\begin{figure}[ht]
\begin{center}
\includegraphics[width=8cm]{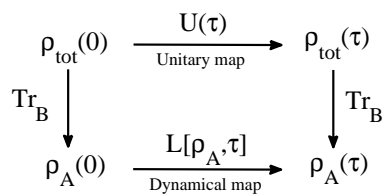}
\caption{Schematic picture of the subsystem dynamics. The full $A+B$ system evolves unitary under the map $U(\tau)$. Tracing over subsystem $B$ gives the state $\rho_A(\tau)$ of subsystem $A$ at the initial ($t=0$) and final ($t=\tau$) times. The dynamical map $L[\rho_A,\tau]$ relates these two states of subsystem $A$. The evolution of $\rho_A(\tau)$ described by the dynamical map is in general not unitary, and even if $\rho_A(0)$ is pure the propagated state $\rho_A(\tau)$ does not need to be pure. }
\label{fig0}
\end{center}
\end{figure}

For completely isolated systems, it is understood that a true thermalization of the full system cannot occur for an initial pure state. Nevertheless, expectation values $\langle\hat{A}_n\rangle$, where $\hat{A}_n$ is any operator that operates in the subspace spanned by particle/subsystem $n$, often display characteristics as if the state $\rho_{mic}^{(n)}$ of this particle/subsystem is a microcanonical one~\cite{eth}. Dividing the full system into two subsystems $A$ and $B$, a sketch of the dynamics is given in Fig.~\ref{fig0}. For more details see~\cite{open,aliki}. The full system state $\rho(0)$ evolves unitarily in time under some operator $\hat{U}(\tau)$. Tracing out the degrees of freedom of subsystem $B$ at time $t=0$ or $t=\tau$ gives the initial and final states for subsystem $A$, $\rho_A(0)$ and $\rho_A(\tau)$. The state $\rho_A(0)$ is related to the state $\rho_A(\tau)$ via a dynamical map~\cite{open,aliki}. When the evolution of subsystem $A$ is Markovian, the dynamical map $L[\rho_A,\tau]$ forms a semi-group~\cite{open,aliki}. The dynamics of subsystem $A$ is in general not unitary. Moreover, when $B$ contains a large number of degrees-of-freedom, any information in $A$ is likely to be distributed over the many degrees-of-freedom of $B$, and $\rho_A(\tau)$ then becomes approximately diagonal in its computational basis. Regardless whether $\rho_A(\tau)$ approaches a microcanonical state or some other diagonal state, system $B$ has a decohering effect on $A$. If $B$ is a closed finite system, we call this decohrence as inherent, while if the decoherence derives from coupling to an infinite reservoir we call it external.  A general believe is that for closed quantum systems, chaos and non-integrability is deeply connected to thermalization, and thereby integrable many-body systems may not thermalize or even result in quasi diagonal states $\rho_A(\tau)$~\cite{ols}. In such systems, recurrences of the coherence of system $A$ often occurs, i.e. the information in system $A$, once lost to $B$, is recovered in $A$ at certain revival times. 

For the photonic many-body systems mentioned above, relatively little has been investigated in terms of inherent decoherence and thermalization. On the contrary, external decoherence and thermalization induced by a surrounding reservoir have been studied in great detail for cavity/circuit QED models~\cite{open,zoller}. There are, however, some exceptions, both theoretical and experimental. In his seminal work~\cite{gea}, J. Gea-Banacloche explained how inherent decoherence may cause peculiar effects in cavity QED systems. Considering the JC model with the field initially in a coherent state with a large amplitude, it was shown how the qubit disentangle from the field at half the revival time, regardless of the initial state of the qubit~\cite{gea}. In the qubit subspace, the evolution is clearly non-unitary since the overlap/angle between any two initial qubit states is not preserved. A physical picture of this phenomenon is given in~\cite{leshouches}. In a couple of experiments performed by the ENS group in Paris, loss of qubit coherence due to interaction with a microwave cavity field was demonstrated~\cite{meter}. A similar idea was later applied in circuit QED, and the qubit dephasing deriving from measurement backaction was studied~\cite{schuster}. In the present work we analyze a plethora of decoherence phenomena, both inherent and external, in a bimodal cavity/circuit QED system. Decoherence originating from; non-unitary subsystem evolution, irreversible coupling to surrounding environments, as well as irreversible coupling to a measurement device are discussed.

In particular, a two-level system (qubit) induces an effective coupling between two quantized photon modes. In the dispersive regime, as is studied here, the qubit mediated tunneling between the two boson modes shows much resemblance with a bosonic Josephson junction, appearing for example in Bose-Einstein condensates in double-well potentials~\cite{smerzi,Jexp} or via coupling of atomic Zeeman levels in a spinor condensate~\cite{intJJ}. The photonic counterpart of the Josephson effect has currently been considered in either coupled cavity systems (where the non-linearity is of a JC photon blockade type)~\cite{photJJ1} or in optomechanical cavity systems (where non-linearity instead stems from a mechanical photon blockade effect)~\cite{photJJ2}. Unlike Refs.~\cite{photJJ1,photJJ2}, the Josephson effect studied in this work is intrinsic and not as in earlier publications where the effect is extrinsic and the Josephson current appears in real space. Moreover, the qubit in our scheme brings out both the boson tunneling as well as the non-linearity. In this respect, our model shares more similarities with the recent works of Ref.~\cite{gatekeep}. The models of~\cite{gatekeep} are, however, both linear. Studying the linear entropy or the qubit Bloch vector, we find that inherent decoherence shows a Gaussian decay signaling non-Markovianity. The non-linearity causes a collapse-revival pattern in the Josephson oscillations, similar to the collapse-revivals found in the JC model~\cite{jc}. In contrast, the linear entropy of the qubit shows a much richer collapse-revival structure containing a series of well established fractional revivals. Measuring the qubit state, and especially the Bloch vector amplitude, provides direct information about the Josephson dynamics. However, the qubit dephasing resulting from non-demolition qubit measurement causes an exponential decay in the oscillations. For non-selective qubit measurements, the Josephson current can be considerably slowed down via the quantum Zeno effect. Somewhat surprising, for a dissipating cavity, the scaled Josephson current does not decay exponentially, but rather Gaussian and it is found that for realistic system parameters around 10 full oscillations survive. The bimodal JC model is integrable and thermalization is not to be expected. Including the counter rotating terms, i.e. going beyond the rotating wave approximation, we consider a weaker sort of thermalization related to the analysis of~\cite{hanggi}. Despite the close similarity with the generic model of Ref.~\cite{hanggi}, we do not encounter thermalization. We further evidence the absence of thermalization by consider long-time state averages as was recently proposed in~\cite{liu}. Neither in this sense does our system thermalize.

The paper continues as follows. In the next section, the model system is introduced and the photonic Josephson effect demonstrated. In the linear regime we present an analytical approximation giving a deeper understanding of the effect. Section~\ref{sec3} proceeds by considering the evolution of the qubit subsystem. Since the full system is so far assumed closed, the field and the qubit subsystems exhibits the same type of reduced evolution. Gaussian decay together with periodic recurrences of the coherence is found. Effects deriving from non-demolition or non-selective qubit measurements are addressed in Sec.~\ref{sec4}, while the following Sec.~\ref{sec5} considers the more complete picture where the cavity has been coupled to a zero temperature photon bath. In Sec.~\ref{sec6}, the prospects of a weak kind of thermalization is analyzed. Finally, concluding remarks are presented in Sec.~\ref{sec7}.

\section{Intrinsic photonic Josephson effect in a bimodal resonator}\label{sec2}
We consider a two-level system, either an atom/ion interacting with an optical Fabry-Perot cavity or a superconducting qubit embedded in a transmission line resonator. We will call the two-level system a qubit and describe it using the Pauli matrices $[\hat{\sigma}_z,\hat{\sigma}^\pm]=\pm2\hat{\sigma}^\pm$ and $\hat{\sigma}_z|\pm\rangle=\pm|\pm\rangle$, where $|\pm\rangle$ are the two internal qubit states. The energy difference between the two qubit states is $\hbar\Omega$. The resonator supports two quantized modes, quasi-resonant with the qubit transition. We will take the two modes as degenerate with frequency $\omega_c$, but point out that the degeneracy is not crucial and could in principle be lifted. Introducing coherent driving of the two modes with amplitude $\eta$, the Hamiltonian becomes ($\hbar=1$)
\begin{equation}\label{ham0}
\begin{array}{lll}
\hat{H} & = & \displaystyle{\omega_c\left(\hat{a}^\dagger\hat{a}+\hat{b}^\dagger\hat{b}\right)+\frac{\Omega}{2}\hat{\sigma}_z}\\ \\
& & +\displaystyle{g_a\left(\hat{a}^\dagger\hat{\sigma}^-+\hat{\sigma}^+\hat{a}\right)
+g_b\left(\hat{b}^\dagger\hat{\sigma}^-+\hat{\sigma}^+\hat{b}\right)}\\ \\
& & +\displaystyle{g_a\left(\hat{a}^\dagger\hat{\sigma}^++\hat{\sigma}^-\hat{a}\right)
+g_b\left(\hat{b}^\dagger\hat{\sigma}^++\hat{\sigma}^-\hat{b}\right)}\\ \\
& & \displaystyle{-i\eta\left(\hat{a}e^{-i\omega_pt}-\hat{a}^\dagger e^{i\omega_pt}\right)-i\eta\left(\hat{b}e^{-i\omega_pt}-\hat{b}^\dagger e^{i\omega_pt}\right)},
\end{array}
\end{equation}
where $\hat{a}$ and $\hat{b}$ ($\hat{a}^\dagger$ and $\hat{b}^\dagger$) are the annihilation (creation) operators of the two modes, $g_{a,b}$ are the qubit-field coupling strengths that have been taken real, and $\omega_p$ is the frequency of the driving field. The third line represents the counter-rotating terms. Turning to a rotating frame with respect to the pumping, and after imposing the rotating-wave approximation (RWA), we have the time-independent Hamiltonian
\begin{equation}\label{ham1}
\begin{array}{lll}
\hat{H}_{RWA} & = & \displaystyle{\delta\left(\hat{a}^\dagger\hat{a}+\hat{b}^\dagger\hat{b}\right)+\frac{\Delta}{2}\hat{\sigma}_z}\\ \\
& & +\displaystyle{g_a\left(\hat{a}^\dagger\hat{\sigma}^-+\hat{\sigma}^+\hat{a}\right)
+g_b\left(\hat{b}^\dagger\hat{\sigma}^-+\hat{\sigma}^+\hat{b}\right)}\\\\
& & \displaystyle{-i\eta\left(\hat{a}-\hat{a}^\dagger\right)-i\eta\left(\hat{b}-\hat{b}^\dagger\right)}.
\end{array}
\end{equation}
Here, $\delta=\omega_c-\omega_p$ and $\Delta=\Omega-\omega_p$ are the cavity-pump and atom-pump detunings respectively. For resonant driving ($\delta=0$) of a cavity without the qubit, energy will be constantly pumped into the two modes. For $\delta\neq0$, on the other hand, the system will reach a coherent steady state with  $\langle\hat{n}_{a,b}\rangle=\eta^2/\delta^2$. 

With $g\equiv\sqrt{g_a^2+g_b^2}$ and $\cos\theta=g_a/g$, we define a new set of boson operators
\begin{equation}
\left[\begin{array}{c}
\hat{A}\\ \hat{B}\end{array}\right]=\left[\begin{array}{cc}
\cos\theta & \sin\theta\\
-\sin\theta & \cos\theta\end{array}\right]\left[\begin{array}{c}
\hat{a}\\ \hat{b}\end{array}\right].
\end{equation}
For $g_a=g_b$, which will be the situation considered throughout this work, the above rotation is a simple Hadamard rotation of the two boson modes. In the lossless case, assuming zero pumping ($\eta=0$), the RWA Hamiltonian is
\begin{equation}
\begin{array}{lll}  
\hat{H}_0 & = & \displaystyle{\delta\left(\hat{A}^\dagger\hat{A}+\hat{B}^\dagger\hat{B}\right)+\frac{\delta}{2}\hat{\sigma}_z}\\ \\
& & +g\sqrt{2}\left(\hat{A}^\dagger\hat{\sigma}^-+\hat{\sigma}^+\hat{A}\right).
\end{array}
\end{equation}
Thus, in the transformed picture the two boson modes are decoupled and the qubit interacts with the $A$-mode according to a JC type. In the large detuning regime, $\Delta\gg g\sqrt{\langle\hat{A}^\dagger\hat{A}\rangle}$, we impose an adiabatic approximation~\cite{jonas0} to find, up to some trivial constants,
\begin{equation}
\hat{H}_0\!=\!\delta\left(\!\hat{A}^\dagger\hat{A}\!+\!\hat{B}^\dagger\hat{B}\!\right)\!+2U\hat{A}^\dagger\hat{A}\hat{\sigma}_z-\frac{4U^2}{\Delta}\!\left(\hat{A}^\dagger\hat{A}\right)^2\!\hat{\sigma}_z+...,
\end{equation}
where $U=g^2/\Delta$. Even though the Hamiltonian is diagonal in the Fock basis of the $A$- and $B$-modes, the original $a$- and $b$-modes still couple in a nontrivial manner since
\begin{equation}
\hat{A}^\dagger\hat{A}=\frac{1}{2}\left(\hat{n}_a+\hat{n}_b+\hat{a}^\dagger\hat{b}+\hat{b}^\dagger\hat{a}\right),
\end{equation}
where the photon numbers $\hat{n}_\alpha=\hat{\alpha}^\dagger\hat{\alpha}$. Neglecting the non-linear terms, the linearized Hamiltonian becomes
\begin{equation}\label{linham}
\hat{H}_0\approx\delta_z\left(\hat{n}_a+\hat{n}_b\right)+U\left(\hat{a}^\dagger\hat{b}+\hat{b}^\dagger\hat{a}\right)\hat{\sigma}_z,
\end{equation}
with the Stark shifted frequencies $\delta_z=\delta+\hat{\sigma}_zU$. The total number of photons, $\hat{N}\equiv\hat{n}_a+\hat{n}_b$, is conserved and the dynamics depends solely on $U$. 

\begin{figure}[ht]
\begin{center}
\includegraphics[width=8cm]{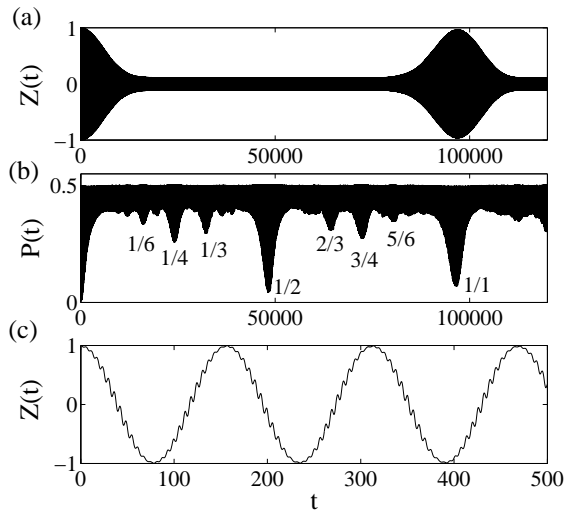}
\caption{Time evolution of the scaled photon inversion Z(t) (a) and (c), and  linear entropy (b) for the Hamiltonian (\ref{ham1}). As an initial state we have a coherent state with amplitude $\alpha=4$ in the $a$-mode, vacuum in the $b$-mode, and the qubit in $(|+\rangle+|-\rangle)/\sqrt{2}$. Due to the non-linearity of the JC model, a clear collapse-revival pattern emerges. The linear entropy displays a more pronounced collapse-revival pattern than the inversion. The dimensionless parameters are $g=1$, $\Delta=50$, $\delta=1$, and $\eta=0.1$. }
\label{fig1}
\end{center}
\end{figure}

With the Hamiltonian $\hat{H}_0$ of Eq.~(\ref{linham}), it is clear that any initial field state $|\psi_a,0\rangle$, where $|\psi_a,\psi_b\rangle$ represents the composite state with the $a$-mode in $|\psi_a\rangle$ and the $b$-mode in $|\psi_b\rangle$, transforms into (up to an overall phase) $|0,\psi_a\rangle$ after a time $t_j=\pi/2U$, regardless of the atomic initial state. This is the analog of intrinsic bosonic Josephson oscillations for photons. In the earlier works on the photonic Josephson effect~\cite{photJJ1,photJJ2} photons tunneled between spatially separated regions, while here they tunnel between two orthogonal cavity modes. To make things more transparent, lets assume the $a$-mode to be initially in a coherent state with amplitude $\alpha$, $\hat{a}|\alpha\rangle=\alpha|\alpha\rangle$, the $b$-mode in vacuum, and the qubit in the superposition state $|\varphi\rangle=(|+\rangle+|-\rangle)/\sqrt{2}$, i.e.
\begin{equation}\label{init}
|\Psi(0)\rangle=|\alpha,0\rangle\frac{1}{\sqrt{2}}\left(|+\rangle+|-\rangle\right).
\end{equation}
In the interaction picture, the time evolved state reads
\begin{equation}\label{timeev1}
\begin{array}{lll}
|\Psi(t)\rangle & = & \displaystyle{\frac{1}{\sqrt{2}}}\Big[|\alpha\left(e^{-i2Ut}-1\right)/2,\alpha\left(e^{-i2Ut}+1\right)/2\rangle|+\rangle\\ \\
& & +|\alpha\left(e^{i2Ut}-1\right)/2,\alpha\left(e^{i2Ut}+1\right)/2\rangle|-\rangle\Big],
\end{array}
\end{equation}
and for the scaled photon inversion defined as
\begin{equation}\label{inv1}
Z(t)\equiv\frac{\langle\hat{n}_a\rangle-\langle\hat{n}_b\rangle}{\langle\hat{n}_a\rangle+\langle\hat{n}_b\rangle}
\end{equation}
we have $Z(t)=\cos(2Ut)$. This result is an ideal case where losses and non-linear contributions have been ignored. In the next sections we will discuss the influences of dissipation and decoherence, while we now consider the consequences of the non-linearity. 

Typically, non-linearity plays an important role in Josephson dynamics. In the bosonic Josephson junction, it gives rise to effects like self-trapping~\cite{smerzi,Jexp} and collapse-revivals~\cite{milburn,rzaz}. Self-trapping appears since the non-linearity can be seen as inducing effective shifts in the two mode frequencies and when these are out of resonance the population swapping is hindered. A collapse in the oscillations derives from the fact that the non-linearity disturbs the equividistant energy spectrum of the linear model (\ref{linham}) and this in turn leads to a destructive interference effect in the dynamics~\cite{robinett}. If at some later time $t_r$, the evolving eigenphases contained in the propagated state return back in phase a revival of the Josephson oscillations occurs. For revivals to be possible, the spectrum should be discrete and quasi-linear, i.e. the non-linearity cannot become to strong. In the language of cavity QED, revivals is a direct proof of the quantized nature of the electromagnetic field~\cite{walther}, while a collapse could, in principle, derive from any fluctuations. We conclude from the fact that our system is governed by JC dynamics, a self-trapping effect should not be found, while collapse-revivals are expected.  

The result for the scaled photon inversion $Z(t)$, obtained from numerically solving the driven two-mode JC model (\ref{ham1}), is presented in Fig.~\ref{fig1}. We truncate the Hilbert space such that $\mathrm{Tr}[\rho(t)]\approx1$, with $\rho(t)$ the full density operator. In (a) we give the inversion for long times, while in (c) we display the short time behavior. The fast oscillating fluctuations in $Z(t)$ seen in (c) derives from the non-zero driving ($\eta\neq0$). The long time dynamics exhibits a typical collapse-revival structure as found in the JC model~\cite{jc,jonasjc,sz}. However, in the present bimodal model the collapse-revival structure is much more complex compared to the regular JC model. In order to demonstrate this we introduce the linear entropy of the field or the qubit~\cite{linent}
\begin{equation}
P_{f,q}(t)=1-\mathrm{Tr}[\rho_{f,q}^2(t)],
\end{equation}
where $\rho_{f,q}=\mathrm{Tr}_{q,f}[\rho]$ is the reduced density operator of the field "$f$" or the qubit "$q$" and $\rho$ is the full system's density operator. Linear entropy is a measure of purity of the state $\rho_{f,q}$, with $P_{f,q}=0$ representing the pure state, and for the special case of a qubit maximum mixedness gives $P_q=1/2$. Whenever the evolution is unitary and the initial state is pure one has $P_f(t)=P_q(t)$~\cite{al}. Therefore, we do not need to distinguish between the two subsystem's linear entropies in this case. For the above example, we plot in Fig.~\ref{fig1} (b) the corresponding linear entropy. This time the collapse-revival structure is much more complex, indicating that rich dynamics is occurring even within the collapse region indicated by (a). The revival-peaks are marked according to standard conventions~\cite{robinett}. In a collapse region, the phase-space distributions are smeared out over the whole accessible part of the phase-space~\cite{dong}. At the 1/2 peak the distribution reforms, however 180 degrees out of phase from its initial position, while at the 1/1 revival it reforms at its original location~\cite{dong}. The remaining fractional revivals represent partial restoration of the phase space distributions, which then form several localized blobs in phase space and not single ones as is the case for the 1/2 and the 1/1 revivals~\cite{dong}. Increasing the non-linearity, i.e. increasing the fraction $g/\Delta$, results in shorter revival periods but also in less pronounced revivals.
 
\section{Field induced decoherence of the qubit}\label{sec3}
Decoherence of a quantum system is seen as loss of information, which is only possible by coupling the system to some other system. If the additional system contains a large number of degrees-of-freedom, the information can be shared in many ways and it takes a long time to restore the information in the initial system. If the number of degrees-of-freedom goes to infinity and the various coupling strengths varies, we expect the information to be forever lost. When we neglect any back-action of the reservoir onto the system we say that the evolution of our system is Markovian~\cite{open}. Markovian dynamics is described by irreversible evolution (i.e. non-unitary), typically leading to exponential loss rates. As the reservoir size is finite or for very structured reservoirs, non-Markovian effects may become important giving rise to recurrences of information, and moreover to non-exponential decay~\cite{open}. In particular, for short times a Markovian decay is linear in time while non-Markovian decay is typically quadratic. In this section we will discuss the inherent decoherence effects the cavity fields induces on the qubit.

In some recent publications, Sokolovski and co-workers considered a quibit coupled to a bosonic Josephson junction~\cite{gatekeep}. A Bose-Einstein condensate trapped in a double-well potential is analyzed. The qubit couples to the barrier separating the two wells, and in this respect its internal state controls the tunneling rate between the condensates. As such, they introduced the name ``quantum gatekeeper model" for this setup. At first, the condensate was seen as a reservoir for the qubit, and the inherent decoherence of the qubit was studied. The qubit as a measuring device for the bosonic Josephson oscillations was then analyzed. In many aspects, the current system shows great resemblance with the gatekeeper model. There are some crucial differences though; the bosonic modes in our model induces coupling between the qubit levels, i.e. our model is intrinsically non-linear, and the photon tunneling is present for both internal qubit states. In principle, as demonstrated in the previous section, transitions within the qubit can be neglected by considering the limit of large detuning. Moreover, by including a second qubit residing all the time in its lower state $|-\rangle$ but still couple identically to the boson modes as the first qubit, it is possible to construct an identical gatekeeper model as in~\cite{gatekeep}. We, however, will keep only a single qubit and allow for qubit transitions and thereby for non-linear effects.

\begin{figure}[ht]
\begin{center}
\includegraphics[width=8cm]{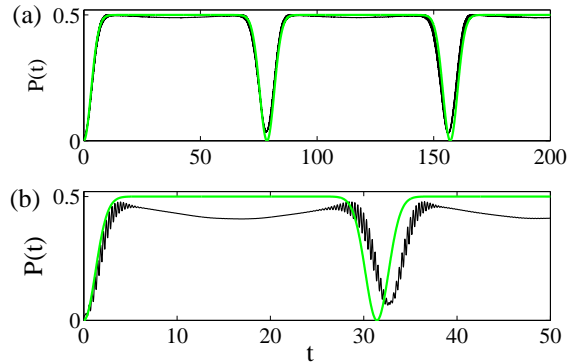}
\caption{(Color online) Comparison of the linear entropy as obtained from the analytical approximate result (\ref{analpur1}) or from numerical diagonalizition of the Hamiltonian (\ref{ham1}) (analytic approximation (green lines) and numerical (black curves)). In (a) the parameters are as in Fig.~\ref{fig1}, i.e. the non-linearities are weak, while in (b) $\Delta=20$ increasing the effects of non-linearities. }
\label{fig2}
\end{center}
\end{figure}

Throughout this work, the two qubit states $|\pm\rangle$ are assumed meta-stable such that dissipation and decoherence can be neglected. In this section we further take any cavity losses to be zero within the time-scale of the experiment. Cavity losses and its effects will be investigated in Sec.~\ref{sec6}. Our initial state will be taken to have the form (\ref{init}). 

The time evolution in the dispersive linear regime is given by Eq.~(\ref{timeev1}), we find the reduced density operator for the qubit
\begin{equation}\label{qubitdens1}
\rho_q(t)=\left[\begin{array}{cc}
\rho_{++} & \rho_{-+} \\
\rho_{+-} & \rho_{--}
\end{array}\right],
\end{equation}
where $\rho_{++}=\rho_{--}=1/2$ and
\begin{equation}
\rho_{-+}=\frac{1}{2}e^{-|\alpha|^2\sin^2(2Ut)}e^{i2|\alpha|^2\sin(2Ut)},
\end{equation}
such that the linear entropy becomes
\begin{equation}\label{analpur1}
P(t)=\frac{1}{2}\left(1-e^{-2|\alpha|^2\sin^2(2Ut)}\right).
\end{equation}
The short time decay of the coherence term $\rho_{-+}$ is quadratic in $t$ as expected for non-Markovian decoherence, and similarly the linear entropy grows quadratically in $t$. In this linear approximate model, at $t=\pi/2U$, the coherence is fully recovered. For unitary evolution with initial pure states, the linear entropy is a good measure of entanglement, and we therefore see that the two fields quickly become entangled with the qubit. For large field amplitudes $|\alpha|$, the two subsystems stays almost completely entangled until they fully disentangle at $t_j=\pi/2U$. Thus, the qubit rapidly loses information (coherence) to the two boson modes, but since the system is finite the information is restored periodically in time. Once the information is completely transfered to the two modes, the qubit is maximally entangled with the cavity fields. Note that the disentanglement occurs when the photonic Josephson junction has performed half an oscillation, i.e. when the initial coherent state in mode $a$ has been transfered to mode $b$. 

The above analytical model is an idealized case where we completely disregarded non-linearities. In Fig.~\ref{fig2}, we therefor compare the analytical results with results obtained from a full numerical diagonalization of the Hamiltonian. As for Fig.~\ref{fig2}, the non-zero pumping induces small rapid oscillations. In the upper plot (a), the parameters are the same as in Fig.~\ref{fig1} and we find that the analytic approximations captures well the overall structures. The oscillations deriving from the pumping is naturally not seen in the analytic results. In the lower plot (b) we decrease the detuning $\Delta$ in order to increase the non-linearity, and as expected the two results begin to disagree. Figure \ref{fig2} gives dynamics at the time-scale of the Josephson oscillations, while in the previous section we studied evolution on the scale of collapse-revivals. Naturally, the linear analytical model of Eq.~(\ref{analpur1}) is not capable of catching the collapse-revival patterns of Fig.~\ref{fig1} (b).

\begin{figure}[ht]
\begin{center}
\includegraphics[width=8cm]{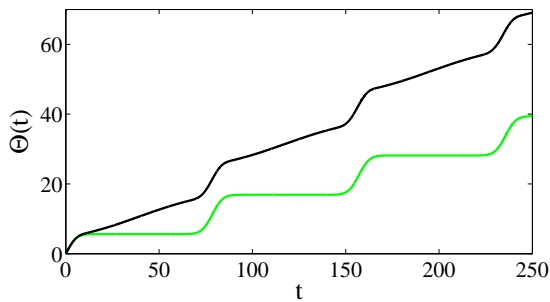}
\caption{(Color online) The integration of the Bloch vector length obtained from the numerical results (black curve) and from the analytical expression~(\ref{analbloch}) (green curve). The parameters are taken the same as in Fig.~\ref{fig2} (a). When integrated, the discrepancy between the analytical and numerical results for the Bloch vector amplitude adds up and the two curves disagree, even though the qualitative structure is the same. }
\label{fig3}
\end{center}
\end{figure}

In the ``gatekeeper model" of~\cite{gatekeep}, the qubit was suggested as a quantum ``meter" measuring the Josephson oscillations of the Bose-Einstein condensate. In that model, if the qubit occupies its upper state $|+\rangle$ Josephson oscillations took place, while if the qubit resided in its lower state $|-\rangle$ oscillations were blocked. Thereby, by measuring the effective amount of time the qubit has spent in the state $|+\rangle$ one knows for how long the oscillations (``the gate") has been turned on (``open"). The situation is different in our case since Josephson oscillations are independent of whether the qubit is in $|+\rangle$ or $|-\rangle$. This, however, does not imply that the qubit does not carry any information about the oscillations. In a Bloch picture, since $\langle\hat{\sigma}_z\rangle\approx0$ the Bloch vector $R(t)\equiv(\langle\hat{\sigma}_x\rangle,\langle\hat{\sigma}_y\rangle,\langle\hat{\sigma}_z\rangle)$ revolve mainly along the equator of the Bloch sphere. Thereby, in our system it should be more practical to measure $\langle\hat{\sigma}_x\rangle$ or $\langle\hat{\sigma}_y\rangle$ instead of $\langle\hat{\sigma}_z\rangle$. Another possibility is to measure the length of the Bloch vector. In our analytic model we have
\begin{equation}
\begin{array}{l}
\displaystyle{\langle\hat{\sigma}_x\rangle=e^{-|\alpha|^2\sin^2(2Ut)}\cos[2|\alpha|^2\sin(2Ut)]},\\ \\
\displaystyle{\langle\hat{\sigma}_y\rangle=e^{-|\alpha|^2\sin^2(2Ut)}\sin[2|\alpha|^2\sin(2Ut)]}
\end{array}
\end{equation}
and for the length of the Bloch vector
\begin{equation}\label{analbloch}
|R(t)|=e^{-|\alpha|^2\sin^2(2Ut)}.
\end{equation}
In the vicinity of the disentanglement times $t_j=n\pi/2U$ ($n=0,\,1,\,2,\,...$), the Bloch vector length is roughly unity. In between the disentanglement, the vector length rapidly shrinks to approximately zero. The area covered by the Bloch vector during one disentanglement period can be approximated by $I=\int\,dt\,e^{-|\alpha|^2(2Ut)^2}=\sqrt{\pi/4U|\alpha|^2}.$ The function $\Theta(t)\equiv\int_0^t\,dt'|R(t')|$ should thereby have a stair-like shape increasing in steps $\sim I$. In Fig.~\ref{fig3} we present the numerical results of $\Theta(t)$ compared to the integration of $|R(t)|$ of Eq.~(\ref{analbloch}). The parameters and initial conditions are the same as for Fig.~\ref{fig2} (a). From the figure it is clear that even in the collapse regions, the two subsystems (fields and qubit) are not fully entangled, which would be reflected in a Bloch vector of zero length. The analytic results (green curve) still captures the general structure of the numerical results (black curve). As a conclusion, by measuring $|R(t)|$, and thereby $\Theta(t)$, the number of Josephson oscillations performed by the cavity fields is obtained. Measuring the qubit dipole movements $\langle\hat{\sigma}_x\rangle$ and $\langle\hat{\sigma}_y\rangle$ is easily obtained by detecting the qubit population inversion after a $\pi/2$ pulse has been applied to the qubit~\cite{haroche}. Any measurement on the qubit will however induce a perturbance, whether it is in the qubit population or in its coherence. While such effects were disregarded from the analysis of Ref~\cite{gatekeep}, we address them in the next section.

\section{Qubit measurement induced decoherenc of the Josephson effect}\label{sec4}
In its original idea~\cite{josephson}, the Josephson effect is a macroscopic quantum phenomenon describing a current across the bulk of two attached superconductors. A common quantum coherence among the constituent electrons is the crucial aspect for a non-zero current. At the mean-field level, the relative phase between the two superconductors are directly coupled to the inversion $Z(t)$~\cite{smerzi}. The question in our system then raises how a qubit measurement will affect the Josephson oscillations. For example, we note that since the Stark-shifts of the qubit states $|\pm\rangle$ depends on the photon number, the back-action noise from measuring the relative phase between the qubit states should manifest itself in the phase coherence of the cavity fields. Apart from dephasing, more drastic outcomes may appear when the measurement is projective. Most well known is the Zeno slowing down caused by repeatedly reprojecting the system state. In this section, we will analyze both the case of pure dephasing and Zeno slowing down. 

\subsection{Measurement induced dephasing}
A qubit measurement relying on for example ionization is, of course, destructive. Once measured, the qubit is ``destroyed". The idea of a quantum non-demolition (QND) measurement is that repeated measurements can be performed on one and the same system, all yielding the same outcome. Let us take a physical quantity characterized by some operator $\hat{A}$, then if $\hat{A}$ commutes with the system Hamiltonian and the measurement of $\hat{A}$ does not induce any back-action noise on $\hat{A}$, a second measurement of $\hat{A}$ will reveal the same value as the first measurement~\cite{qnd}. Even though the measurement does not have any back-action effect on $\hat{A}$, according to quantum mechanics it will induce noise in the conjugate variable of $\hat{A}$. In cavity QED, in order to measure the photon number of the quantized cavity mode, QND measurements has been thoroughly analyzed both theoretically~\cite{qndtheo} and experimentally~\cite{qndexp1}. Similar QND measurements have been carried out also in circuit QED~\cite{qndexp2}. Our goal is not, however, to measure the number of photons contained in the resonator, but to measure the state of the qubit. For superconducting qubits, such QND measurements have recently been demonstrated~\cite{qndqubit}. The picture we have in mind is that the the qubit level-population is detected non-destructively, i.e. measuring $\hat{\sigma}_z$. Since neither $\hat{\sigma}_x$ nor $\hat{\sigma}_y$ commutes with $\hat{\sigma}_z$, the measurement induces a qubit dephasing~\cite{qndbackaction}. In the bosonic Josephson junction, decoherence has been shown to suppress the oscillations~\cite{rzaz,decJJ}. For Bose-Einstein condensates in double-well potentials, the loss of coherence (phase diffusion) is mainly due to atom-atom interactions~\cite{phasdiff}. The dephasing we consider in this section has a different nature as it affects the anchilla system and not the boson modes directly.

\begin{figure}[ht]
\begin{center}
\includegraphics[width=8cm]{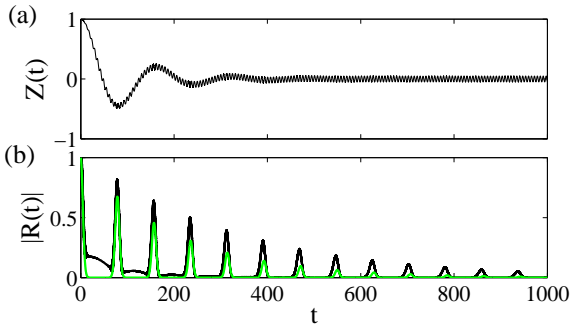}
\caption{(Color online) The scaled photonic inversion $Z(t)$ of Eq.~(\ref{inv1}) and the amplitude of the Bloch vector $|R(t)|$, as obtained from numerical integration of the master equation (\ref{master1}) (black curves). The green curve shows the comparison with the analytical approximation (\ref{analR}) for the Bloch vector amplitude. Both the Josephson oscillations and the Bloch vector decay exponential due to the qubit measurements. The decoherence rate $\gamma=0.005$, and the remaining parameters are as in Fig.~\ref{fig1}.}
\label{fig4}
\end{center}
\end{figure}

We assume this QND measurement to be an irreversible process that can be modeled by a master equation~\cite{open}, i.e. the measurement is continuous and far from a traditional von Neumann measurement. As no dissipation occurs, we write down the equation-of-motion for the full density operator as
\begin{equation}\label{master1}
\begin{array}{lll}
\displaystyle{\frac{\partial\rho(t)}{\partial t}} & = & -i[\rho(t),\hat{H}_{RWA}]\\ \\
& & +\gamma\left(2\hat{\sigma}_x\rho(t)\hat{\sigma}_x-\hat{\sigma}_x^2\rho(t)-\rho(t)\hat{\sigma}_x^2\right),
\end{array}
\end{equation}
where $\gamma$ is the effective decoherence rate induced by the measurements. Letting $\gamma=0$ we find identical evolution as in Figs.~\ref{fig1} and~\ref{fig2}. Once $\gamma\neq0$, the off-diagonal terms $\rho_{-+}$ and $\rho_{+-}$ of the atomic density operator will typically show an exponential decay~\cite{open}.

We solve the master equation (\ref{master1}) by simple Newton integration, valid to order $\mathcal{O}(dt^2)$ in the time-step $dt$. However, in its simplest form, the first part describing regular Hamiltonian dynamics becomes non-unitary and the method is greatly unstable. Therefore, the first unitary evolution is taken into account in all orders of $dt$, while the irreversible evolution is performed to order $\mathcal{O}(dt^2)$. As long as $dt$ is small this should be justified, and to be fully sure we have verified convergence of our results by varying $dt$. 

In Fig.~\ref{fig4}, the photonic inversion (a) together with the Bloch vector amplitude (b) are displayed. It is seen that the Josephson oscillations decay exponential, indicating that the noise induced by the qubit measurement propagates to the fields and thereby demolish the macroscopic coherence needed for the oscillations. As pointed out in the introduction, this exponential decay signals an irreversible loss of coherence. The Bloch vector amplitude shows an interplay between the non-Markovian inherent decoherence resulting from coupling to the two cavity modes and Markovian external decoherence due to the qubit measurements. The exponential decay rate is set by $\gamma$, and in the most simple approximation we get
\begin{equation}\label{analR}
|R(t)|=e^{-|\alpha|^2\sin^2(2Ut)-\gamma t}.
\end{equation}
From Fig.~\ref{fig4} (b) we note that the analytical approximation~(\ref{analR}) (green curve) slightly overestimates the exponential decay. A large $\gamma$ implies that information about the qubit, and thereby the boson modes, is rapidly leaked from the system to the ``observer/meter". If the measurement is perfect, i.e. a complete knowledge of the system is obtained, all the coherence shared between the two subsystems gets lost, and the qubit disentangles from the two modes. There is hence a trade-off between the information subtracted and the survival of the Josephson oscillations. We note that using the qubit as a meter was recently utilized in a circuit scheme measuring the quantum state of a mechanical (rather than an optical) resonator~\cite{connell}.

\subsection{Quantum Zeno effect}
Repeatedly measuring a system affects the effective evolution of the system. The extreme situation of freezing the evolution by constantly projecting the system state back onto its initial state is the idea behind the quantum Zeno effect~\cite{zeno1}. In reality, no measurement is expected to be instantaneous resulting in a ``dead-time'' after each measurement. In this respect an ideal Zeno effect is unlikely~\cite{zeno1}. This, however, does not prevent a slowing down effect, which was recently demonstrated within cavity QED~\cite{zeno2}. We briefly discussed the evolution of a cavity mode under coherent external pumping. For an empty mode, the field growth is initially quadratic in time. Under a Zeno slowing down, this becomes linear in time~\cite{zeno2}. A quadratic growth hinges on coherently adding up the injected field, while repeated QND measurements destroys the field phase. The injected field can then be seen as making a random walk in phase space and thereby the linear growth. We will not consider the field build up due to external pumping, but instead consider the growth of the initially empty cavity mode deriving from photon tunneling between the two modes. We have seen that such growth is as well quadratic for small times $t$. In order to separate it from external pumping we pick $\eta=0$. 

The approach of the previous subsection is not capable of describing a Zeno slowing down since we pictured a constant loss of information, and not repeated phase disturbances. We will therefore instead assume repeated non-selective qubit measurements~\cite{nonsel}. At times $t_n=n\tau$ ($n=1,\,2,\,...$) the qubit population is measured but the result is not recorded. The correlations between the fields and the qubit is taken to be fully demolished by the measurement, hence
\begin{equation}
\rho(t)\rightarrow\rho_f(t)\rho_q(t).
\end{equation}
In between measurements, the system evolves unitarily as $\hat{U}_\tau=\exp\left(-i\hat{H}_{RWA}\tau\right)$. Since the measurements are taken to be instantaneous, the full evolution operator becomes
\begin{equation}
\hat{U}(t,\tau)=\Big[\hat{M}\hat{U}_\tau\Big]^k,
\end{equation}
where $\hat{M}$ is the field-qubit disentanglement operator characterizing the measurement, and $k$ is the largest integer such that $0\leq t-k\tau\leq1$. If $\tau$ is small, the coherent build up of the field is frequently interrupted and we expect a slow and linear field increase.  

\begin{figure}[ht]
\begin{center}
\includegraphics[width=8cm]{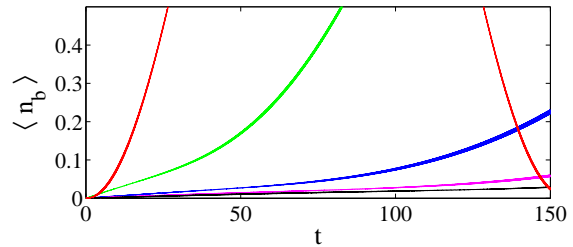}
\caption{(Color online) Evolution of the field intensity $\langle\hat{n}_b\rangle$ under repeated non-selective qubit measurements. The different curves correspond to different measurement frequencies; red curve $\tau=\infty$, green curve $\tau=1/100$, blue curve $\tau=1/500$, pink curve $\tau=1/1000$, and black curve $\tau=1/1500$. The dimensionless parameters are, $g=1$, $\Delta=50$, $\delta=1$, $\eta=0$, and $\alpha=2$. }
\label{fig5b}
\end{center}
\end{figure}

The predicted linear rise of the field amplitude is numerically demonstrated in Fig.~\ref{fig5b} for various values of $\tau$. For no measurements, $\tau=\infty$, the field grows quadratically from time $t=0$. When $\tau$ is decreased we find a regime of linear growth which passes over into a quadratic regime. The smaller $\tau$, the smaller is the incline and the more inhibited is the quadratic evolution. The figure is a typical example of a Zeno slowing down~\cite{zeno2,zeno3}. 

For a composite system $A+B$, it has been demonstrated that repeated projective measurements on  subsystem $A$ render a purification of subsystem $B$~\cite{pur1}. Letting the $\rho(0)=\rho_A(0)\rho_B(0)$, with $\rho_A(0)=|\psi_A(0)\rangle\langle\psi_A(0)|$, be the initial state, the projective measurements divides the unitary evolution at $t_n=n\tau$ ($n=1,\,2,\,...$) according to
\begin{equation}
\rho(t)\rightarrow\rho_B(t)|\psi_A(0)\rangle\langle\psi_A(0)|.
\end{equation}   
In the case of non-selective measurements a disentanglement between the two subsystems appears, while for projective measurements in addition to disentanglement the $A$ system is as well reset to its initial state. Regardless of the initial state $\rho_B(0)$, the effective evolution will translate into a pure state. This is the idea behind measurement controlled evolution~\cite{pur1,pur2}. In the present system, we have verified that this mechanism indeed works. Various initial states like coherent and thermal field states have been considered, and in every case the purity of the field subsystem approaches unity after sufficiently many projective qubit measurements. The specific purified field state differs between various initial field states. In certain situations the fields approach Fock states, but this seems not to be a generic feature.

\section{Reservoir induced decoherence}\label{sec5}

So far we have considered decoherence as either inherent or imposed via qubit measurements. We now proceed by analyzing the effects deriving from coupling the photon modes to a zero temperature reservoir. The two qubit states $|\pm\rangle$ are still taken to be metastable and losses are therefore solely in the boson fields. As for the case of qubit measurements, we model the irreversible coupling by a master equation~\cite{zoller}
\begin{equation}\label{master2}
\begin{array}{lll}
\displaystyle{\frac{\partial\rho(t)}{\partial t}} & = & -i[\rho(t),\hat{H}_{RWA}]\\ \\
& & +\kappa\left(2\hat{a}\rho(t)\hat{a}^\dagger-\hat{n}_a\rho(t)-\rho(t)\hat{n}_a\right)\\ \\
& & +\kappa\left(2\hat{b}\rho(t)\hat{b}^\dagger-\hat{n}_b\rho(t)-\rho(t)\hat{n}_b\right).
\end{array}
\end{equation}
Here, $\kappa$ is the common decay rate of the two modes. The above equation is commonly used for a zero temperature photon reservoir, the generalization to non-zero temperature typically implies two extra Lindblad terms that are as well proportional to $\kappa$. In the case of strong qubit-field coupling, it has been shown that such a master equation does not in general decay into the correct canonical distribution~\cite{canjc}. However, at zero temperature as considered here, the long time behavior of the above master equation is physically relevant~\cite{canjc}. In contrast to Eq.~(\ref{master1}) which describes decoherence, the above equation (\ref{master2}) results in dissipation of cavity photons to the reservoirs. As will be seen, pure dephasing and dissipation have different influences on the Josephson oscillations. Direct dephasing of the boson modes, without dissipation, is obtained by replacing $\hat{a}$ and $\hat{a}^\dagger$ by $\hat{n}_a$ (and similarly for the $b$ mode) in the two Linblad terms.  

\begin{figure}[ht]
\begin{center}
\includegraphics[width=8cm]{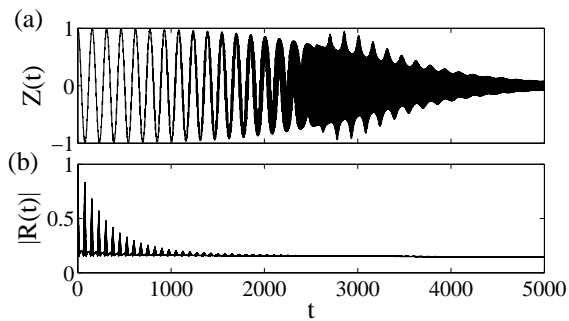}
\caption{Evolution of the scaled photon inversion $Z(t)$ (a) and the Bloch vector amplitude $|R(t)|$ (b) in the presence of photon dissipation. The Bloch vector amplitude displays a combination of Gaussian and exponential decay. The scaled Josephson oscillations, however, survives much longer. The parameters are as in Fig.~\ref{fig1}, but with $\alpha=3$ and $\kappa=0.0012g$ corresponding to the experimental decay rate of Ref.~\cite{wal}. }
\label{fig6}
\end{center}
\end{figure}

Exactly like in the previous sections, we can find an analytical solution of Eq.~(\ref{master2}) in the linear dispersive regime~\cite{klimov}. In the long time limit and letting $\delta=0$, the atomic density operator becomes
\begin{equation}
\rho_{at}(t)=\frac{1}{2}\left[\begin{array}{cc}
1 & \rho_{+-}\\
\rho_{-+} & 1
\end{array}\right],
\end{equation}
where
\begin{equation}
\rho_{+-}=\tilde{p}e^{-\kappa\eta^2\left(\frac{U}{\kappa^2+U^2}\right)^2t}
\end{equation}
and $\tilde{p}$ is a complex time-independent factor of order unity~\cite{klimov}. The form of the field density operator reads
\begin{equation}
\rho_f=\frac{1}{2}\left[|a_+,b_+\rangle\langle a_+,b_+|+|a_-,b_,\rangle\langle a_-,b_-|\right],
\end{equation}
where the explicit expressions for the coherent amplitudes $a_\pm(t)$ and $b_\pm(t)$ in the most general case are rather complex, see Ref.~\cite{klimov}. They become, however, simple in the limit of no pumping, $\eta=0$, 
\begin{equation}
\begin{array}{l}
\displaystyle{\langle\hat{n}_a\rangle=\frac{\alpha^2}{2}e^{-2\kappa t}\left[1+\cos(2Ut)\right]},\\ \\
\displaystyle{\langle\hat{n}_b\rangle=\frac{\alpha^2}{2}e^{-2\kappa t}\left[1-\cos(2Ut)\right]}.
\end{array}
\end{equation}
It follows that the inversion $Z(t)=\cos(2Ut)$ as without losses. In general though, with $\eta\neq0$, the expression for the inversion becomes very complicated. Nonetheless, for this case we note that generally $\langle\hat{n}_a\rangle\neq\langle\hat{n}_b\rangle$ in the limit $t\rightarrow\infty$. This derives from the fact that the two mode were initially differently populated. As a consequence of this, the photonic inversion does not approach zero; $\lim_{t\rightarrow\infty}Z(t)\neq0$. Moreover, as long as $\delta\neq0$, the initial decay of $Z(t)$ is not exponential, which is in contrast to pure dephasing. We will also see that non-linearities beyond the above analytical model play an important role. 

As before, our numerical investigation goes beyond the linear regime of the analytical consideration. We solve the master equation (\ref{master2}) with the same procedure as in the previous section. For the photon decay rate we take an experimentally relevant value of $\kappa=0.0012g$~\cite{wal}. The numerical results are shown in Fig.~\ref{fig6}. The rest of the parameters are the same as in the example of Fig.~\ref{fig1}, except that the amplitude of the initial coherent state $\alpha=3$ (instead of $\alpha=4$ as in Fig.~\ref{fig1}). As in the case of pure dephasing, the Bloch vector amplitude displays an interplay between Gaussian decay and recurrences and exponential decay. Contrary to the example of dephasing, the amplitude does not approach zero in the long time limit. This does not, however, imply that the qubit is entangled with the boson modes. Since the evolution is not unitary, the linear entropy (nor the Bloch vector amplitude) is a good measure of entanglement. In fact, the off-diagonal elements of both $\rho_{at}$ and $\rho_f$ vanish in the long time limit.

It is interesting to note that the behavior of the system with photon dissipation is very different from the case with dephasing. Dephasing induced an exponential decay of the photon numbers $\langle\hat{n}_a\rangle$ and $\langle\hat{n}_b\rangle$ such that $\langle\hat{n}_a\rangle=\langle\hat{n}_b\rangle$ for large times. The same kind of exponential decay was seen in the photon inversion $Z(t)$ (see Fig.~\ref{fig4} (a)). Dissipation causes as well an exponential decay of $\langle\hat{n}_a\rangle$ and $\langle\hat{n}_b\rangle$ towards their steady state values, but $Z(t)$, on the other hand, does not decay exponentially. At first, the scaled Josephson current is not greatly affected by the dissipation. After $\sim6$ full Josephson oscillations, the scaled current begin to decay, however with something more reminiscent of Gaussian decay. As the amplitudes of the two modes decrease, the rapid oscillations deriving from the non-zero pumping becomes more pronounced, and after around 20 full oscillations the pumping induced fluctuations washes out the Josephson oscillations and a new structure in $Z(t)$ emerges. This additional inference pattern is not seen in the regime where the Josephson oscillations are still well resolved. It is important to point out, even though the Josephson oscillations may survive relatively long, it should be remembered that the photon amplitudes $\langle\hat{n}_a\rangle$ and $\langle\hat{n}_b\rangle$ can become very small putting limitations on photon detection. Today, on the other hand, very small field amplitudes ($\langle\hat{n}\rangle\ll1$) can be measured~\cite{haroche}.

In order to confirm the differences between dephasing and dissipation of the boson modes (Figs~\ref{fig4} and \ref{fig6}), we have numerically solved the dephasing master equation, i.e. replacing $\hat{a}$ and $\hat{a}^\dagger$ by $\hat{n}_a$ and replacing $\hat{b}$ and $\hat{b}^\dagger$ by $\hat{n}_b$ in the Linblad terms. This time, the inversion $Z(t)$ decays exponentially in a similar way as in Fig.~\ref{fig4} (a). 

Figure~\ref{fig6} has been calculated using physically relevant parameters and we thereby conclude that the intrinsic photonic Josephson effect should indeed be realizable. However, only around 10 Josephson oscillations survive. On the other hand, also for the intrinsic and extrinsic bosonic Josephson effect studied in Bose-Einstein condensates is typically only detected during a few oscillations~\cite{Jexp,Jexp2}. Our numerical analysis does not allow us to go to larger photon numbers, but increasing the initial amplitude $\alpha$ may prolong the oscillation time. It should be noted though, that increasing the number of photons while keeping the detuning $\Delta$ fixed increases the non-linearity, which might lead to undesired effects. Stronger pumping will help in terms of preserving a measurable number of photons in the two modes, but it will as well imply a bigger impact of rapid oscillations on top of the Josephson oscillations. Optimizing the system parameters might be tricky, but it is nonetheless clear that the example of Fig.~\ref{fig6} is probably not at all the most optimal one in terms of a sustainable Josephson effect. 

\section{Quantum thermalization?}\label{sec6}
Thermalization in quantum systems is normally attributed to dissipation and external decoherence~\cite{open,zoller}. Assuming a quantum system $S$ to be irreversibly coupled to a thermal reservoir with some characteristic temperature $T$, and denoting the state of the system as $\rho_s$, by waiting long enough the state of the system normally becomes diagonal and on the form
\begin{equation}\label{gibbs}
\rho_s\propto e^{-\beta\hat{H}_s},
\end{equation}
where $\hat{H}_s$ is the system Hamiltonian and $\beta=k_BT$ is the Boltzmann's constant time temperature $T$. The reduction of the off-diagonal coherence terms of $\rho_s$ is an outcome of the dissipation and decoherence induced by the reservoir. On the other hand, assuming $\rho_s$ to be initially pure and the system to be closed, it follows that $\rho_s$ will remain pure, i.e. $\mathrm{Tr}[\rho_s^2(t)]=1$. However, this does not forbid the state $\rho_A$ of subsystem $A$ of the total system $S$ to become diagonal. We saw examples of temporary subsystem diagonalization in the previous sections. This, of course, is an effect of the inherent decoherence enforced upon subsystem $A$. Whether a recurrence of the off-diagonal terms appears is a highly legitimate question. More precisely, one wishes to understand if, why, and how $\rho_A$ attains an asymptotic diagonal distribution that can be modeled by a microcanonical ensemble. When thermalized, the assymptotics of any expectation value of some operator $\hat{A}$ acting on the subsystem $A$ can be obtained from statistical mechanical arguments. This is a topic of current intense research~\cite{eth}. A rule of thumb is that a chaotic system thermalizes in the above sense. The present model, in its closed counterpart, Eq.~(\ref{ham1}), is not chaotic and contains only a few degrees of freedom, and one thereby do not expect thermalization. In fact, we know that for relatively localized initial states ($\langle\Delta\hat{n}\rangle\sim\langle\hat{n}\rangle$), the JC model possesses a series of collapse-revivals~\cite{jc}, and we will not have any thermalization. Things may become more interesting if we go beyond the RWA and include the counter rotating terms to our Hamiltonian. The model is then believed to be non-integrable~\cite{jonasjc,irrjc}. Even so, within the range of our numerics, generic thermalization is not found in our model. Instead, we will consider an even weaker type of thermalization that was recently discussed by Ponomarev {\it et al.} in Ref.~\cite{hanggi}. Their model does not put any constrains on integrability nor on a large number of degrees-of-freedom. In this respect it is more related to the system at hand.

\begin{figure}[ht]
\begin{center}
\includegraphics[width=8cm]{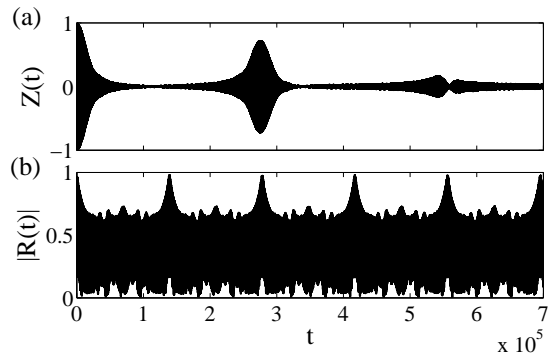}
\caption{Evolution of the scaled photon inversion $Z(t)$ (a) and the Bloch vector amplitude $|R(t)|$ (b) for an initial thermal state in mode $a$ and vacuum in $b$. The inversion shows a Gaussian collapse, and at later times revivals build up indicating absence of thermalization. The Bloch vector amplitude collapses and revive at half the inversion revival and also at the revival of the inversion. In between these revivals a series of fractional revivals are seen. The dimensionless parameters are $\omega_c=54.3g$, $\Omega=150g$, and $\bar{n}_a=5$. }
\label{fig7}
\end{center}
\end{figure}

The Hamiltonian considered in~\cite{hanggi} has the form
\begin{equation}\label{hang}
\hat{H}_P=\hat{H}_A+\hat{H}_B+\hat{H}_{int},
\end{equation}
where $\hat{H}_A$ and $\hat{H}_B$ act in subsystems $A$ and $B$ respectively, and they are assumed identical $\hat{H}_A=\hat{H}_B$, and furthermore the interaction is taken on the form $\hat{H}_{int}=\hat{Y}_A\otimes\hat{Y}_B$ where $\hat{Y}_A=\hat{Y}_B$ and they too act in the respective subsystems. Thus far, the model is rather generic. However, one more constrain is assumed, letting $E_n$ be the $n$'th eigenvalue of $\hat{H}_P$ they assumed $E_n-E_m=E_s-E_w$ if and only if $s=n$ and $w=m$. Within this model, Ponomarev {\it et al.} showed that for an initial state $\rho(0)=\rho_A(0)\otimes\rho_B(0)$, with $\rho_{A,B}(0)$ thermal states characterized by the temperature $T_{A}$ and $T_B$ respectively, the asymptotic states $\rho_A(\infty)=\rho_B(\infty)$. Hence, regardless of the two temperatures $T_A$ and $T_B$, the system thermalizes to states with a common temperature. It was further demonstrated that this property is not restricted to initial thermal states, but holds for more general diagonal initial states $\rho_A(0)$ and $\rho_B(0)$. 

As a closed system~(\ref{hang}), we will in this section neglect any irreversible coupling to reservoirs. Thinking of subsystem $A$ and $B$ as our two boson modes, it is clear that our model does not fulfill the above constrains. First of all, the qubit adds extra degrees-of-freedom, and secondly, not even in the linear regime (i.e. the importance of the qubit degrees-of-freedom has been omitted) does our Hamiltonian possesses the desired form (\ref{hang}). Furthermore, within the RWA, the linearized model has a spectrum that does not obey $E_n-E_m\neq E_s-E_w$ for different $(n,m)$ and $(s,w)$. This condition can, however, be met by including the counter rotating terms, in other words go beyond the RWA. To have a time-independent model we do not consider external pumping, and our Hamiltonian to be studied in this section is then
\begin{equation}\label{hamth}
\begin{array}{lll}
\hat{H} & = & \displaystyle{\omega_c\left(\hat{a}^\dagger\hat{a}+\hat{b}^\dagger\hat{b}\right)+\frac{\Omega}{2}\hat{\sigma}_z}\\ \\
& & +\displaystyle{g\left[\left(\hat{a}^\dagger+\hat{a}\right)
+\left(\hat{b}^\dagger+\hat{b}\right)\right]\hat{\sigma}_x}.
\end{array}
\end{equation}     
We know that the qubit induces an effective coupling of the two modes, and that the counter rotating terms brings about an irregularity of the spectrum. Therefore, it feels natural to expect our non-RWA system to thermalize in the sense of Ref.~\cite{hanggi}. 

As our initial state we let the qubit be in the same pure state as earlier, mode $b$ in vacuum, and mode $a$ in a thermal state
\begin{equation}
\rho_A(t=0)=\sum_{n=0}^\infty\frac{\bar{n}_a^n}{(\bar{n}_a+1)^{n+1}}|n\rangle\langle n|,
\end{equation}
where $\bar{n}_a$ determines the number of thermal photons, i.e. $\langle\hat{n}_a\rangle=\bar{n}_a$. At zero temperature, $T=0$, we naturally have $\bar{n}_a=0$. Note that beyond the RWA, the number of excitations is not conserved. If we assume that the energy exchange between the fields and the qubit is small within the large detuning regime, thermalization then implies
\begin{equation}\label{therm}
\rho_A(\infty)=\rho_B(\infty)=\sum_{n=0}^\infty\frac{\bar{n}^n}{(\bar{n}+1)^{n+1}}|n\rangle\langle n|, 
\end{equation}
with $\bar{n}=\bar{n}_a/2$.

Our numerical results, presented in Fig.~\ref{fig7}, demonstrates a Gaussian collapse of the photon inversion into a state with a thermal distribution. However, after some time the Josephson oscillations sets off again and the system revive. The two collapsed field states showed identical thermal like photon distributions, but the off diagonal coherence terms of the states are not exactly zero which causes the revival. In the JC model, the revivals become more fuzzy with time and after very long times the revivals begin to overlap causing super-revivals. The well resolved collapse-revival structure of Fig.~\ref{fig7} (a) seems to survive for extremely long times, and within our numerical tests we have surprisingly not seen any long time decay of the revivals. The amplitude of the revivals varies, but every now and then a revival appears where $Z(t)$ oscillates approximately between -1 and 1. Furthermore, we have checked our findings for different initial populations $\bar{n}_a$ and $\Delta$. However, if $\Delta$ is taken too small the non-linearity is very strong and the whole collapse-revival structure smears out. The Bloch vector amplitude displayed in Fig.~\ref{fig7} (b) reproduces a similar collapse-revival structure as was found in Fig.~\ref{fig1} (b), i.e. consisting of a series of full and fractional revivals. Thus, once more does the qubit possesses a more complex dynamics than the Josephson current. The increased qubit purity at half the revival time is reminiscent of that encountered in the regular JC model~\cite{gea}. Not shown in Fig.~\ref{fig7} is the field purity. Despite the fact that the initial state is mixed and one would expect thermalization, we have found that the field purity displays a similar structure as the Bloch vector amplitude. Thus, the structure of the field purity contradicts as well thermalization. We should point out that we have also verified similar results as the ones above when the qubit is an initial maximally mixed state, $\rho_{at}(t=0)=(|+\rangle\langle+|+|-\rangle\langle-|)/2$.  

\begin{figure}[ht]
\begin{center}
\includegraphics[width=8cm]{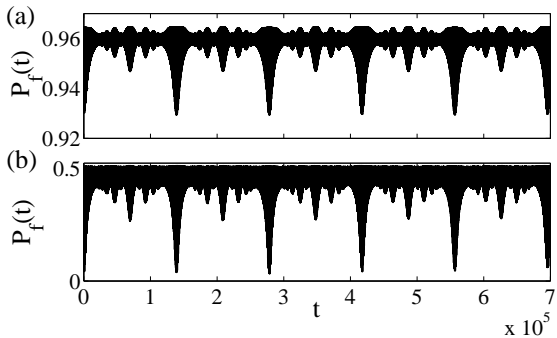}
\caption{Evolution of the field linear entropy $P_f(t)$ for an initial mixed field state of mode $a$~(\ref{cohmix}) (a), and a coherent state with the same photon distribution as the mixed state (b). The $b$-mode is initially in vacuum and the qubit as in previous examples. The general structure is similar in both cases, but the initial mixed state stays largely mixed even at the revivals in contrast to the initial pure state which almost perfectly regains its purity at the half and full revivals. The dimensionless parameters are as in Fig.~\ref{fig7}, except the initial mean number of photons $\bar{n}_a=16$. }
\label{fig8}
\end{center}
\end{figure}

Figure~\ref{fig7} hints that our system does not thermalize. To explore this further, we consider instead an initial diagonal state
\begin{equation}\label{cohmix}
\rho_A(t=0)=\sum_{n=0}^\infty\frac{|\alpha|^{2n}}{n!}|n\rangle\langle n|.
\end{equation} 
Thus, despite being diagonal the photon distribution is Poissonian as for a coherent state. In Fig.~\ref{fig8} we display the corresponding field linear entropy $P_f(t)$ (a). This is compared with the field linear entropy with an initial pure state $\rho_A(t=0)=|\alpha\rangle\langle\alpha|$. The general structure is the same between the two cases, but it is clear that even at the well resolved revivals the purity of the initially mixed state is not increased considerably. Nonetheless, it clearly demonstrate the absence of thermalization. 

Finally we explore a different definition of thermalization which was discussed in Ref.~\cite{liu}. With the model Hamiltonian~(\ref{hamth}), the time-averaged reduced density operators
\begin{equation}  
\bar{\rho}_\alpha=\lim_{T\rightarrow\infty}\frac{1}{T}\int_0^Tdt\,\rho_\alpha(t),\hspace{1cm}\alpha=A,\,B,
\end{equation}
where analyzed. If $\bar{\rho}_\alpha$ becomes diagonal, the system clearly has thermalized. We do not show our results here, but we have verified for different initial states and parameters that the time-averaged reduced density operators do not diagonalize. Thus, once again we have found absence of thermalization.

\section{Conclusions}\label{sec7}
We have presented a most simple model that simulates the photonic counterpart of a bosonic Josephson junction. Opposed to earlier studies of the photonic Josephson effect, our model describes the internal effect. In the dispersive regime a qubit mediates an effective coupling between two photon modes of either an optical or a transmission line resonator. After introducing the general model we demonstrated the Josephson effect both within a linear analytic model and using full numerical diagonalization. Our numerical results shows the presence of collapse-revivals originating from the intrinsic qubit-field non-linearity. 

We continued by addressing several aspects that arises from decoherence in one or another way. To start with, we considered inherent decoherence of the qubit stemming from interaction with the quantized boson modes. In such a Markovian analysis we found a rapid Gaussian decay of the qubit purity, at quarter of a Josephson oscillation period the qubit has lost almost entirely its coherence. At half an oscillation, the coherence is essentially completely regained. Our setup serves as well as a realization of the "quantum gatekeeper model´´ discussed in~\cite{gatekeep}. As in Ref.~\cite{gatekeep}, regarding a qubit coupled to a boson Josephson junction formed in a Bose-Einstein condensate~\cite{gatekeep}, we also examine the prospects of using the qubit as a quantum meter for the Josephson oscillations. We show that the Josephson current can be measured by integrating the length of the Bloch vector. 

What was not put forward in~\cite{gatekeep} is the backaction a qubit measurement has on the Josephson current itself. As a first example, we mimicked the irreversible qubit measurement within a master equation approach. The induced dephasing of the qubit rapidly propagates to the boson modes which leads to an exponential decay of the oscillations. The more information subtracted from the qubit, the stronger is the demolishing effect on the Josephson oscillations. This fact demonstrates the macroscopic nature of the Josephson oscillations as a collective quantum many-body phenomenon. As a second example, we considered non-selective or projective measurements, and such interrupted unitary evolution was shown to render a quantum Zeno slowing down of the Josephson oscillations. 

The following section looked into effects deriving from cavity dissipation. Again, this was modeled utilizing a master equation. The important result coming out from this analysis was the fact that even when realistic experimental parameters was considered, the Josephson oscillations survived for around 10 full oscillations. This is indeed comparable, or even longer, than what has been observed in either Bose-Einstein condensates trapped in double-well potentials~\cite{Jexp}, or in spinor condensates with Raman coupled Zeeman levels~\cite{Jexp2}. A somewhat surprising finding was that photon dissipation did not render an exponential decay of the scaled Josephson current, but instead we found a Gaussian decay. This is in contrast to the exponential decay of the amplitudes of the two modes, and also to the exponential decay of the scaled current when exposed to dephasing.

We finished our investigations by considering thermalization in a weak sense. That is, we did not study equilibration into microcanonical ensembles where the expectation values $\langle\hat{A}\rangle$ can be predicted by the corresponding ensembles~\cite{eth}, but thermalization of two interacting thermal quantum systems according to~\cite{hanggi}. Our model fulfills most of the constrains for such thermalization, but it was demonstrated that despite this we do not find thermalization. Lack of the sort of thermalization as discussed in~\cite{liu} was also numerically verified. 

Throughout we assumed the two boson modes to be degenerate. Our results would still be valid even if we drop this constrain. Then, however, should it be noted that the Josephson effect is detuned and hence not all photons will be swapped between the two modes. In circuit QED, which utilizes transmission line resonators, there is in general only a single mode for a given frequency. However, in optical Fabry-Perot cavities there are typically two orthogonally polarized degenerate modes. Noteworthy is that bimodal cavity QED experiments have already been demonstrated~\cite{bimodexp}.

\begin{acknowledgments}
The author acknowledges 
financial support from the Swedish Research Council.
\end{acknowledgments}

\end{document}